\documentclass[]{spie}  
\usepackage[dvips]{graphicx}

\title{Wide-field Imaging Interferometry Testbed II: Implementation, Performance, and Plans}

\author{S. A. Rinehart\supit{a}, Bradley J. Frey\supit{b}, David T. Leisawitz\supit{b}, Douglas B. Leviton\supit{b},\\ Anthony J. Martino\supit{b}, William L. Maynard\supit{b}, Lee G. Mundy\supit{b,c}, Stacy H. Teng\supit{c},\\ Xiaolei Zhang\supit{b,d}
\skiplinehalf
\supit{a}National Research Council Associate, NASA Goddard Space Flight Center.\\
\supit{b}NASA Goddard Space Flight Center.\\
\supit{c}University of Maryland, College Park.\\
\supit{d}Science Systems and Applications Inc.
}

\authorinfo{Further author information: (Send correspondence to S.A.R.)\\S.A.R.: E-mail: rinehart@rosette.gsfc.nasa.gov, Telephone: 1 301 286 3124, Address: NASA -- Goddard Space Flight Center, Mail Code 631, Greenbelt, MD 20771.}

\begin{document} 
\maketitle 

\begin{abstract}
The Wide-Field Imaging Interferometry Testbed
(WIIT) will provide valuable information for the development of
space-based interferometers.  This laboratory instrument operates
at optical wavelengths and provides the ability to test operational
algorithms and techniques for data reduction of interferometric
data.  Here we present some details of the system design and
implementation, discuss the overall performance of the system to
date, and present our plans for future development of WIIT.  In
order to make best use of the interferometric data obtained with
this system, it is critical to limit uncertainties within the
system and to accurately understand possible sources of error.  The
WIIT design addresses these criteria through a number of ancillary
systems.  The use of redundant metrology systems is one of the most
important features of WIIT, and provides knowledge of the delay
line position to better than 10 nm.  A light power detector is used
to monitor the brightness of our light sources to ensure that small
fluctuations in brightness do not affect overall performance.  We
have placed temperature sensors on critical components of the
instrument, and on the optical table, in order to assess
environmental effects on the system.  The use of these
systems provides us with estimates of the overall system
uncertainty, and allows an overall characterization of the results
to date.  These estimates allow us to proceed forward with WIIT,
adding rotation stages for 2-D interferometry.  In addition, they
suggest possible avenues for system improvement.  The possibility
exists to place WIIT inside an environmentally controlled chamber
within the Diffraction Grating Evaluation Facility (DGEF) at
Goddard in order to provide maximum control over environmental
conditions.  Funding for WIIT is provided by NASA Headquarters
through the ROSS/SARA Program and by the Goddard Space Flight
Center through the IR\&D Program.
\end{abstract}



\section{INTRODUCTION}
\label{sect:intro}

The Wide-Field Imaging Interferometry Testbed (WIIT) is a lab-based
instrument for testing the concepts considered for interferometers in
space.  It will be used to test algorithms and data acquisition
schemes which will be used with future interferometry missions and
will provide information which will impact the design and operation of
these future missions.  The overview and aims of WIIT are discussed
further in an accompanying paper$^1$ and in references therein.

In this paper, we focus on some of the details of the design and
operation of WIIT.  In Section~\ref{sect:design}, we talk about some
of the aspects of the design and configuration of the system.  In
Section~\ref{sect:tests}, a number of tests used to test the ability
and limits of the system are discussed. 
Finally, in Section~\ref{sect:future}, we
mention approaches to improving the system and expanding its
capabilities.

\section{DESIGN}
\label{sect:design}

The overall design of WIIT has been challenging because of the
stringent requirements of interferometry.  These requirements have
impacted the mechanical design of the system and the need for accurate
metrology.  In addition, they constrain our optical design and the
electronics implementation (both in terms of hardware and software).
An overview of the design is provided in Leisawitz, et
al. (2002)$^{1}$.  Here we concentrate on some of the details of the
design and implementation.  In particular, we focus on electronic and
software implementation and the overall operation of the system.
Details of the mechanical design, system metrology, and optical design
are included in an accompanying paper$^{2}$.

\subsection{Electronic Design and Software}

   \begin{figure} \begin{center} \begin{tabular}{c}
   \includegraphics[height=7cm]{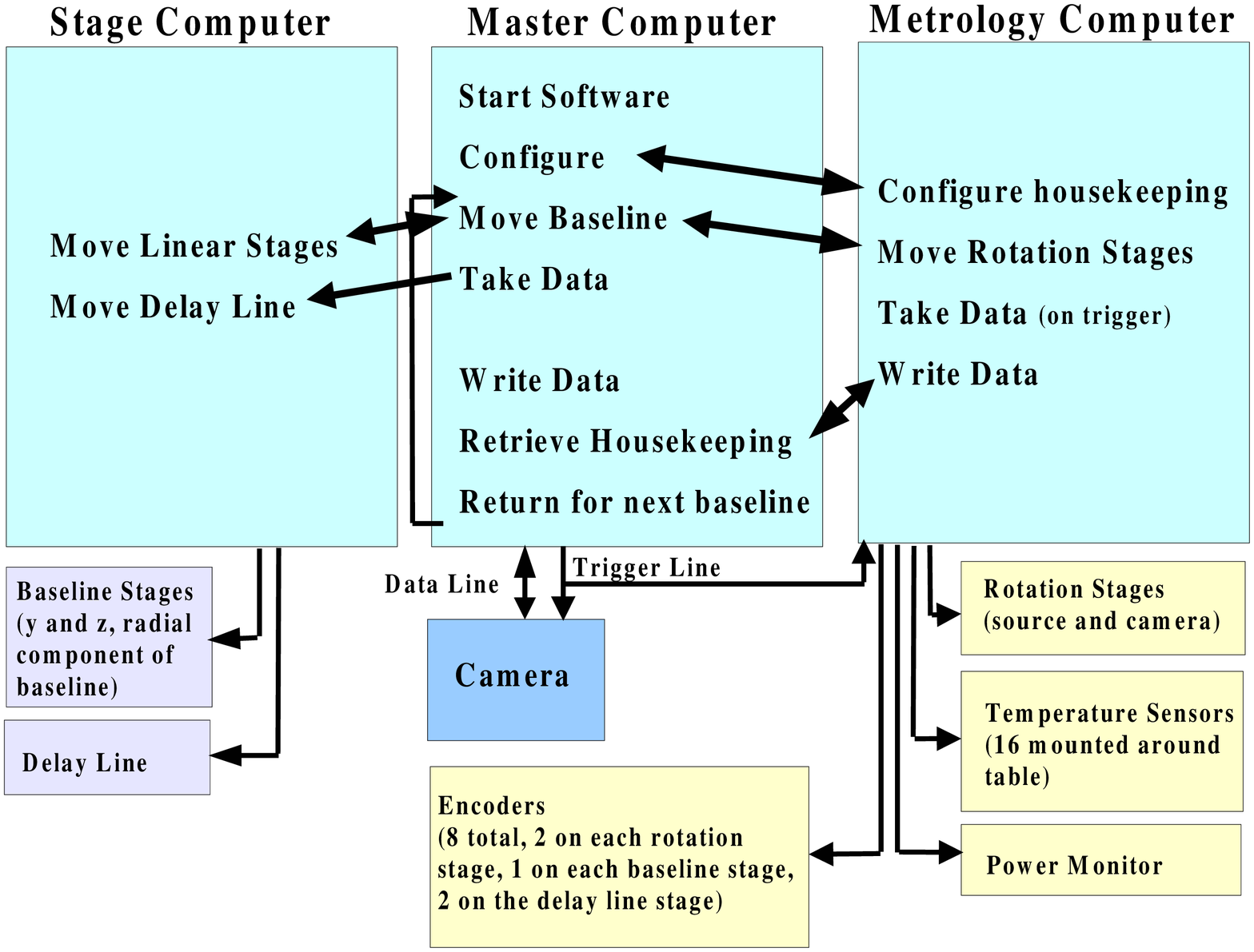}
   \end{tabular}
   \end{center}
   \caption[example] 
   { \label{fig:elecblock} A block diagram of the electronics systems and the interactions between them.  The Master PC controls the other two computers in the system, which in turn control individual hardware elements in the system.}
   \end{figure} 

   \begin{figure} \begin{center} \begin{tabular}{c}
   \includegraphics[height=7cm]{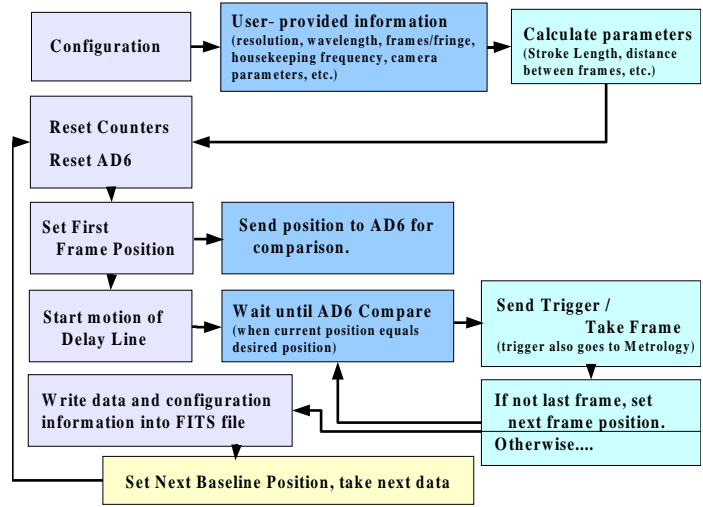}
   \end{tabular}
   \end{center}
   \caption[example] 
   { \label{fig:software} The software operation of the system, as shown in pseudo-code.  The program configures the system, then initializes the individual components, and then begins data acquisition.  When the full scan is complete, the data and configuration information are written to a FITS file, and the baseline is modified for a new scan.}
   \end{figure} 

The electronics of WIIT are, in general, relatively uncomplicated.
There are three computers used in the system; the Master computer, the
Metrology computer, and the Stage computer.  The Metrology computer
controls the two rotation stages, housekeeping systems, and reads the
encoders used to monitor position of the stages$^2$.  The Stage
computer controls the motion of the three linear stages (two baseline stages
and the delay line stage) via software provided by Aerotech.  The
Master computer coordinates the actions of the three computers and is
used for acquisition of images from the camera.  A block diagram of
this can be seen in Figure~1.

The three computers used in the WIIT system interact in several ways
to simplify the process of data acquisition and system operation.  The
Master PC runs a program written in LabView to coordinate the
computers and provide data acquisition.  Figure~2 provides a
pseudo-code diagram of how this process works.  In words, the program
queries the user for parameters of the run, including such information
as central wavelength, desired spectral resolution, and number of
frames per fringe.  Based upon this information, the program
calculates the total stroke length necessary and the delay line
distance between successive frames.  The AD6 is then reset, so that we
can accurately measure relative position.  This also ensures that the
AD6 is in the proper mode to conform with the expectations of the
software.

During this initialization process, the LabView prgram also
coordinates with the Metrology PC to begin monitoring of environmental
parameters.  The LabView program creates a configuration file (based
upon user-accessable parameters).  The metrology software resident on
the Metrology PC reads this file and configures the housekeeping
systems (encoders, temperature sensors, power monitoring) accordingly.
The metrology program then enters a data acquisition standby mode, and
sends a handshake signal back to the Master PC to indicate that it is
prepared.

Data acquisition commences on Master PC at this point.  Based upon the
parameters calculated during initialization, the position value for
the first data frame is set into a buffer of the AD6.  The Master PC
then sends a signal to the Stage PC to start the motion of the delay
line stage.  When the position value in the AD6 equals that of the
value in the buffer, it sends a signal (COMPARE) to the LabView
program.  The LabView program sends a trigger signal to the camera and
records the data from the frame.  The electronic trigger generated by
this procedure is split out to both the camera and to the Metrology
PC.  This trigger signals the Metrology PC to take housekeeping data.
This provides the distinct advantage of simultaneity of data streams,
and avoids the problem of interpolation of two related, but
differently time-sampled, data sets. The program then sets the
position value for the next frame, and loops repeating this procedure
for each value of delay line position calculated.

When the last frame has been taken (i.e. when $N_{frames}$ is the same
as calculated during initialization), the data is written as a FITS
file.  Simultaneously, a signal is sent to the Metrology PC that data
acquisition is complete.  The Metrology PC stops housekeeping data
acquisition, and writes out seperate files for the power monitoring,
temperature data, and encoder readings.  When these files have been
written, the Metrology PC handshakes with the Master PC and the Master
PC retrieves these files.  Post-run, these data care incorporated into
a single FITS table for simplicity and ease of use.

\subsection{Operation and Data Acquisition}

As described above, the software/electronic design of WIIT makes
operation and data acquisition very straightforward for the user, with
an intuitive GUI for the user.  However, there are certain other
aspects of the system that must be addressed when taking data and some
features which merit mention.

While our system is very stable over long timescales, it is still
deemed important to verify that the system is properly aligned and
operating as expected prior to taking interferometric
data\footnote{Because the system has been undergoing a number of
improvements over time, and installation of new elements can cause
small alignment shifts, this has been especially necessary to date.
With completion of the system, this should become less necessary.}.
When preparing to take data, we confirm that the two arms of the
interferometer are aligned by alternately blocking the individual arms
and observing the position of a point source on the camera array.  If
the two spots are not aligned, they can be quickly and easily brought
into alignment by adjusting micrometer control of the beam combiner.
Next, we confirm that the brightness level of the source is suitable.
Because we operate with a white light source using several different
bandpass filters and a laser, light levels can vary significantly.  We
therefore choose an appropriate neutral density filter and adjust
camera parameters (integration time, gain) to achieve maximum signal
without saturating the detector.  At the same time, we adjust the
camera settings to only display and record a subset of the CCD array
choosing only to record data for the field-of-view of interest.  For
example, when observing a point source, we typically use an 8x8 subarray of the
camera, choosing the subarray so that the point source is centered in
the frame.  This provides access to all of the relevant data while
keeping the size of data files written to a minimum.

   \begin{figure} \begin{center} \begin{tabular}{c}
   \includegraphics[height=9cm]{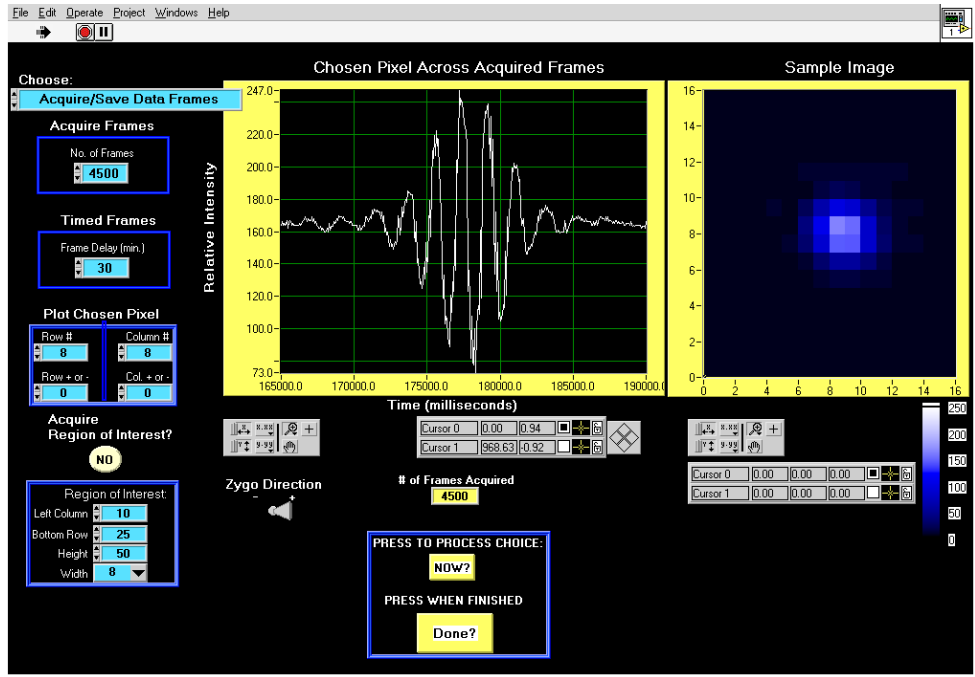}
   \end{tabular}
   \end{center}
   \caption[example] 
   { \label{fig:guiss} A screenshot of the Graphical User Interface (GUI) used on the Master PC.  The plot has been zoomed within the GUI to show clearly white light fringes.}
   \end{figure} 

When taking data, the GUI provides several forms of useful feedback (see Figure~3).
For system debugging purposes, it provides a running frame count,
which has been valuable for understanding our triggered acquisition
mode and limitations thereof.  In addition, at the end of each run,
the GUI displays the image of the last frame and a plot of the
brightest pixel versus frame number.  The latter can be manipulated to
show different pixel values, or to show average pixel values inside a
rectangular box.  This is very useful for individual data runs (single
baseline measurements) and especially for checking system health, as
it allows us to quickly estimate the location of ZPD, fringe contrast,
and the overall quality of the data.  There are also quick look
features built into the software which runs on the Metrology PC
controlling the temperature sensors, power monitor, and encoders.
This is discussed further in a companion paper$^2$.

After data acquisition, the data is written onto a hot-swappable hard
drive on the Master PC.  For small data files, the data can be
transferred via ftp or scp commands (or even by email).  For
wide-field of view data, however, a single scan at one baseline, using
a 100$\times$100 subarray of the detector, produces several hundred Mb
of data, depending upon the resolution desired.  Multiply this by the
number of baselines used, and the volume of data becomes enormous (see
Leisawitz, et al.$^1$), making these simple data transfer methods
unacceptable.  Therefore, we have multiple hot-swappable hard drives;
one in the WIIT lab, one on the analysis computer at Goddard, and one
on an analysis/development computer at the University of Maryland.
When we produce large sets of data, the hard drive is physically
transported to the computer where it will be reduced and switched the
the drive there.  This maximizes fidelity of the data when copied onto
other computers, reduces concerns about bandwidth issues between
different sites, and provides us a method for making system and
software backups with triple redundancy.

\section{TESTING AND SYSTEM BEHAVIOR} \label{sect:tests}

\subsection{System Focus}

   \begin{figure}
   \begin{center}
   \begin{tabular}{c}
   \includegraphics[height=7cm]{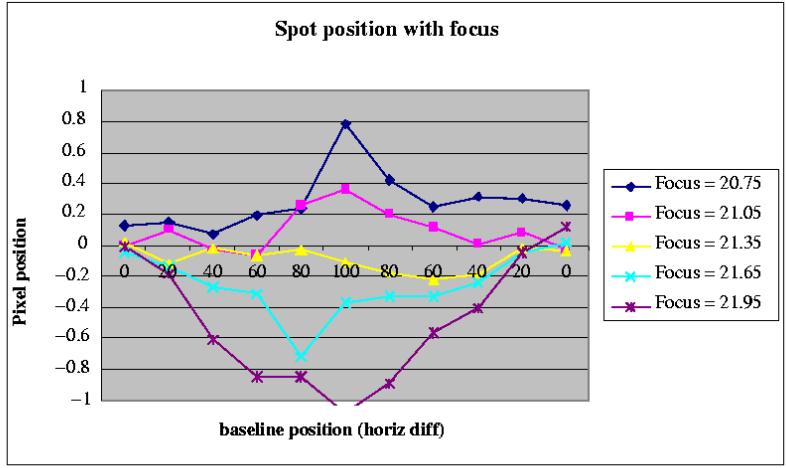}
   \end{tabular}
   \end{center}
   \caption[example] 
   { \label{fig:focus} This plot shows the relative offset of the spots produced by the two arms of the interferometer as a function of baseline distance (in mm per arm).  For clarity, the x axis is plotted from 0 mm (shortest baseline) to 100 mm and back to 0 mm, as this produces an easily interpretable curve.  We see that small deviations from focus can lead to relative shifts of over a pixel, while for our best focus position, the spots produced by the two arms remain coaligned to better than 0.2 pixels.}
   \end{figure}

One of the major requirements on the setup of WIIT is very accurate
system alignment and focus.  It is critical, in order to maintain
fringe quality, that the collimated beam entering the receiver mirrors
is parallel, and that all of the optical components of the
interferometer itself are well-aligned.  The alignment of the system,
and the basic procedures used to focus the system are discussed in a
companion paper$^2$, but here we discuss the final focussing
procedure.

If we were dealing with a single aperture, lack of focus could be
observed directly through an out-of-focus spot on the detector
produced by the divergent/convergent rays.  In the case of an
interferometer, the individual receiver mirrors are much smaller and
cannot resolve focus issues in this way except for very large
deviations.  However, the divergence of the rays does manifest in an
easily measurable way; if the system is not in focus, then movement of
the baseline mirrors results in a measurable change in the position of
an imaged point source.  This is obviously problematic for an
interferometer, as when the two arms of the interferometer move to
different baselines, the images that each of the two arms produce
shift by similar amounts in {\em opposite} directions.  Obviously, it
would be difficult to produce interferograms if the two images do not
coincide to relatively high accuracy.  In Figure~\ref{fig:focus}, we
show the differential position of the spots produced by each arm as a
function of baseline position (where 0 is the shortest baseline and
100 is the longest baseline).  We can see that for even small changes
in focus position, the two spots begin to seperate at different
baselines (and that they seperate in opposite ways on either side of
focus, as one would expect).  Further, we find that at our best
position, the two spots remain coaligned to within 0.2 pixels (peak to
peak).  This method has allowed us to achieve optimal focus for the
system to within about 0.15 mm, where this number is limited by the
uncertainty of repeatability for individual pixel positions (we have
found that the individual pixel positions have 1$\sigma$ uncertainties
of 0.1 pixels) and by the straightness of the rail upon which the
baseline stages travel.

\subsection{Servo Jitter}

   \begin{figure}
   \begin{center}
   \begin{tabular}{c}
   \includegraphics[height=7cm]{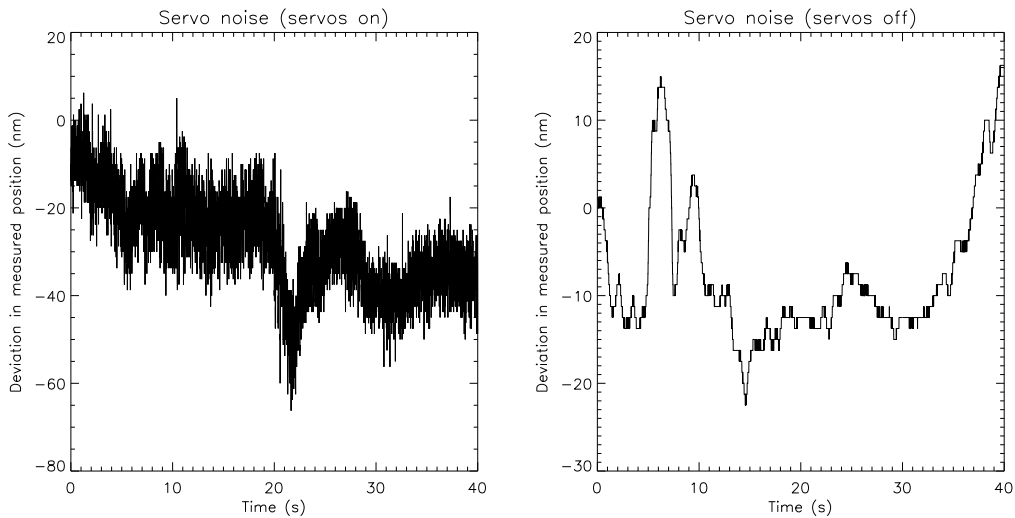}
   \end{tabular}
   \end{center}
   \caption[example] 
   { \label{fig:jitter} Two plots of the position of the delay line stage, as measured with a distance measuring interferometer (Zygo).  The plot on the left shows the measured position (in nm) as a function of time while the stage servos are operating but while the stage is stationary.  The plot on the right shows the same, but with the servos deactivated.  From these plots, it is clear that the servos produce a signficant amount of high-frequency noise in the system.}
   \end{figure}

After we first acquired fringes with WIIT, we observed that the fringe
quality was lower than expected.  We found that one of the significant
causes of this is the behavior of the delay line stage.  There are two
issues regarding the delay line stage which were cause for concern;
first, there is servo position error in the stage, which leads to
periodic velocity fluctuations while the stage is in motion.  This in
turn causes non-uniform spacing of data points when running an
open-loop acquisition.  Second, there is stage jitter caused by the
servos.  This was explored by using a distance measuring
interferometer (Zygo) to measure the position of the delay line stage
while the stage was stationary, but with the servos in operation.  In
Figure~5, we show the plots of delay line position as a function of
time, first with the servos on then with them off.  Environmental
effects are clearly present in both plots, but the high frequency
noise is unique to the data taken with servos operating.  This
high-frequency noise has a 1$\sigma$ value of $\sim 8$nm.
Fortunately, we had anticipated the latter as a possible problem, and
therefore had planned to use a Zygo interferometer to mitigate it by
closing the data loop.  The Zygo measures the position of the delay
line directly, and sends a quadrature signal to a converter (an AD6
purchased from U.S. Digital) attached to the parallel port of the
Master PC.  The AD6 then tracks the position of the delay line stage
and produces ``COMPARE'' pulses when the value reaches the position
set by the Master PC.  It is this pulse which is used to trigger
camera acquisition and housekeeping operation.  This reduces the
problems of non-uniform velocity of the stage and stage jitter from
the data, by linking each frame directly to an exact position of the
delay line stage.  However, while eliminating this problem has
significantly improved the data produced by WIIT, the data of Figure~5
suggest another issue.

\subsection{Turbulence Testing}

After implementing the closed-loop data acquisition, we found that
there were still some noise issues which were unaccounted for.  The
postulate was that we were witnessing noise due to environmental
effects such as temperature changes in the room and air turbulence.

   \begin{figure}
   \begin{center}
   \begin{tabular}{c}
   \includegraphics[height=18cm]{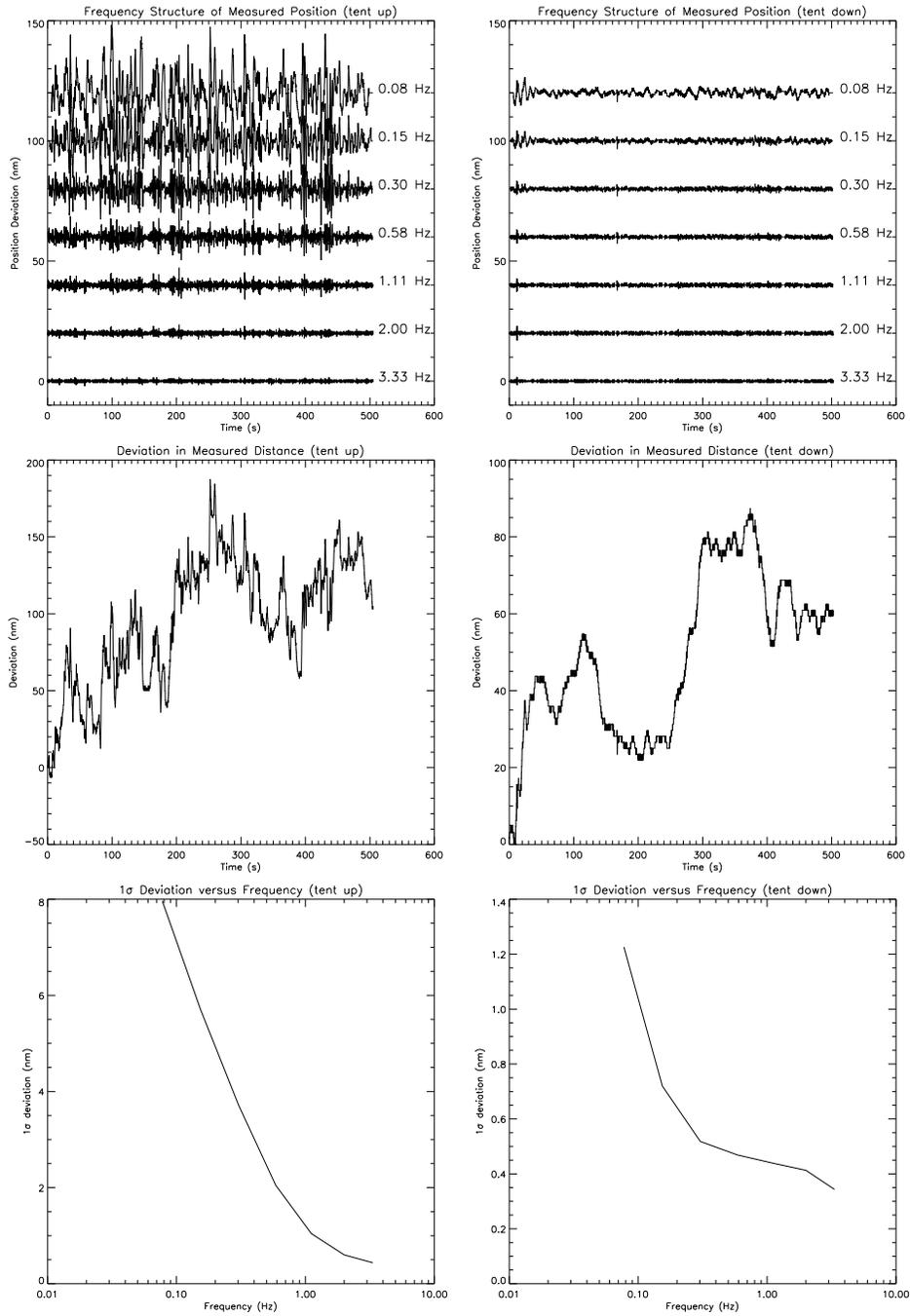}
   \end{tabular}
   \end{center}
   \caption[example] 
   { \label{fig:turb1} This set of plots show the effects of turbulence in several different ways.  The set of three plots on the left were taken without the tent around the system, while the set of three on the right made use of the tent to reduce environmental effects.  The top plots show the data with smoothed data subtracted, in order to estimate the amount of noise at different frequencies.  The middle plots are the raw position data, and the bottom plot shows the 1$\sigma$ deviations as a function of frequency, using the plots shown in the top row.  We see the noise increases going to lower frequencies, and that the effects are greatly reduced by using the tent for environmental isolation.  All of the data taken for these plots was taken with fixed mirrors and a total path length of 20 cm.}
   \end{figure}

Using the Zygo interferometer, we directly examined the effects of
turbulence on our system.  We used the Zygo in conjunction with flat
mirrors mounted directly to the optical table to measure variations in
the ``length'' of the path.  We found that the ``length'' of the path
deviated significantly over time, as expected.  This is shown clearly
in Figure~\ref{fig:turb1}.  In this figure, we used a relatively short
path length (20 cm) in order to observe the effects of the
environment.  We found that the noise was not significant at
frequencies above about 1 Hz, but that at lower frequencies, the noise
was significant.  This result was also borne out at longer path
lengths (up to 200 cm), which also showed that the noise due to
environmental effects is close to linearly correlated with path
length.  We also found that the system was acoustically sensitive and
was vulnerable to motion near the apparatus.  The table was therefore
enclosed in a ``tent'' comprised of black plastic of an aluminum frame
and the test sequence was repeated.

   \begin{figure}
   \begin{center}
   \begin{tabular}{c}
   \includegraphics[height=6cm]{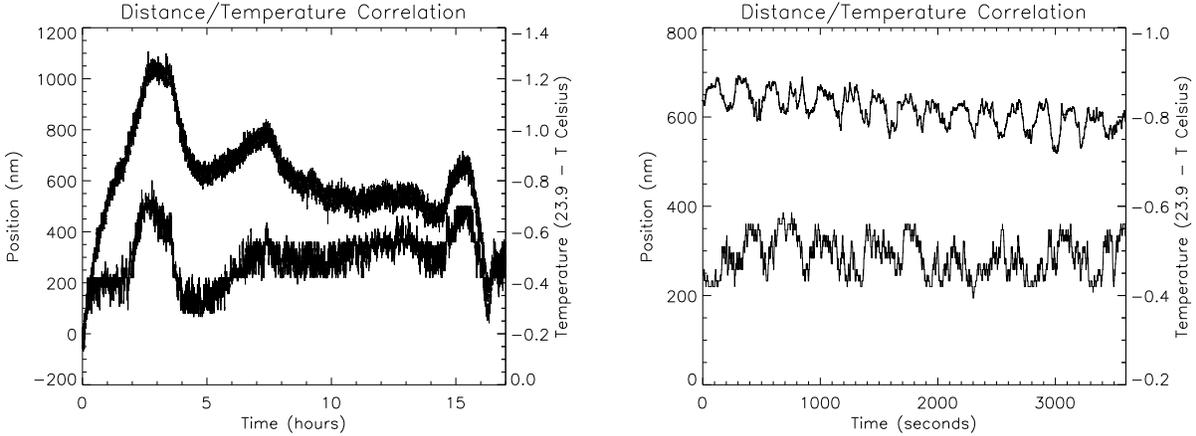}
   \end{tabular}
   \end{center}
   \caption[example] 
   { \label{fig:turbtemp} Two plots which show the correlation between measured position and temperature.  On the left is a plot which shows the position (upper line) as a function of time, as well as the measured temperature at one of the sensors (bottom line).  We see that the overall structure is very similar.  The plot on the right shows a similar plot, for a single hour of the data.}
   \end{figure}

We found that the tent significantly reduced noise at all frequencies.
Comparision of the plots shown in Figure~6 show clearly the
significance of this reduction.  However, we still see evidence for
variation in the measured distance.  To further explore this behavior,
we ran the turbulence test overnight, while taking temperature data
with the sensors located on the optical table.  We found (see
Figure~7) that long-term variations in position correlated well with
room air temperature.  This is not unexpected, as temperature changes
of air cause small variations in the index of refraction of air, which
directly impacts the length of the path as measured by the Zygo
interferometer.  By directly measuring this correlation, we can use
temperature monitoring data to correct for variations in the measured
position of the delay line stage due to thermal effects.  Note that
the shift in position is about three times larger than expected for
the measured temperature shift (in dry air), but that this is
accounted for by thermal expansion/contraction of the table and by the
effects of water vapor in the air.  Figure~7, however, illustrates
another problem.  If turbulence and thermal effects have an impact on
the laser of our Zygo metrology system for these short path lengths,
they also could cause relative phase shifts between the two arms of
the interferometer.  To further improve, we are undertaking two system
improvements.  First, we will install another level of environmental
baffling, by enclosing the interferometer in a hard box.  Second, we
have ordered a two-channel digital Zygo interferometer.  This will be
used to measure the path length of the two arms of the interferometer
seperately, so that we can trigger camera frames based upon the
relative path length difference of the two arms.  When this system is
implemented, we will have substantially reduced environmental effects
on the system.  The tent also has the side benefit of reducing stray
light within the system.  Without the tent and with roomlights on we
found a small additional noise contribution ($\sim$2\%).  With room
lights off, this contribution becomes very small, but obviously
impairs other work within the lab.  The addition of the tent allows
light isolation for WIIT while still allowing full use of the rest of
the lab.

In Figure~8, we can see how the two improvements to date, the addition
of closed-loop operation and the tent for environmental shielding,
have improved operation.  The improvement between open-loop and
closed-loop data is very clear upon examination of this figure.  The
improvement due to the environmental shielding is less clear, although
there is certainly some evidence that the data is less noisy.  The
primary advantage of the environmental shielding, however, is at very
low frequencies, which will have the most significant impact on longer
data acquisition runs.  Since these fringes were taken over a short
period of time (a few minutes), the effects of the lowest frequency
variations cannot be seen.  The two improvements discussed here should
significantly reduce noise in the data.

   \begin{figure}
   \begin{center}
   \begin{tabular}{c}
   \includegraphics[height=7cm]{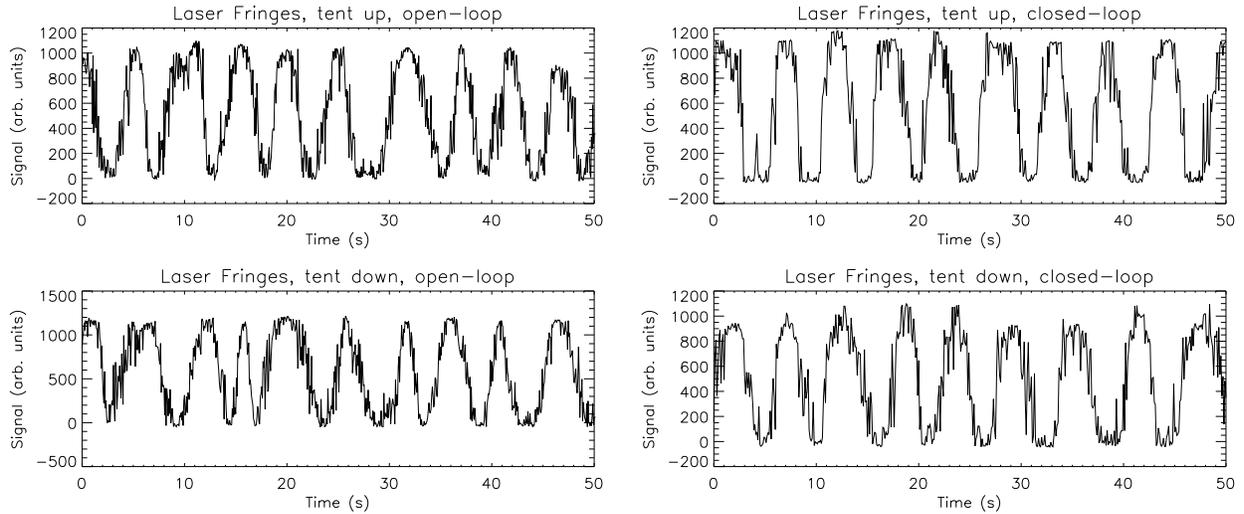}
   \end{tabular}
   \end{center}
   \caption[example] 
   { \label{fig:turbtemp} These four plots show laser fringes under different conditions.  The left pair were taken open loop, the right pair used the DMI for closed-loop operation.  The top pair were taken with the tent up, and the bottom pair with the tent down to reduce environmental impact.  We see that the fringes in the closed-loop data are much more evenly spaced than in the open-loop data.  Further, we see that there is some noise reduction with the tent down than with the tent up.}
   \end{figure}

\subsection{System Stability}

One of the major concerns for WIIT is the long-term stability of the
system.  Is the point of zero path difference (ZPD) vulnerable to
wander over time?  How often does the system need to be checked to
verify that the two arms are aligned and that everything is
functioning properly?  Since, in its final incarnation, WIIT will
operate entirely under software control for several days at a time,
ensuring that the entire system is stable over this time period is
critical.  To answer these questions, we have developed a series of
``system health tests'' which are used to monitor the long-term
behavior of the system.  These tests are enumerated in the Table~1.

\begin{table}
\caption{The list of system health tests which are used to monitor the long-term stability of the system.}
\label{tab:fonts}
\begin{center}       
\begin{tabular}{|l|l|l|} 
\hline
\rule[-1ex]{0pt}{3.5ex} & Test & Purpose\\
\hline
\hline
\rule[-1ex]{0pt}{3.5ex} 1 & Dark Frames & Dark Current and Read Noise\\
\hline
\rule[-1ex]{0pt}{3.5ex} 2 & Flat Fields & Uniformity of response, calibration\\
\hline
\rule[-1ex]{0pt}{3.5ex} 3 & Spot position with baseline & Quality of Focus\\
\hline
\rule[-1ex]{0pt}{3.5ex} 4 & Power monitor with baseline & Uniformity of Illumination\\
\hline
\rule[-1ex]{0pt}{3.5ex} 5 & White Light Stability (arm blocked) & Image quality and stability\\
\hline
\rule[-1ex]{0pt}{3.5ex} 6 & Laser Stability & Image quality and stability\\
\hline 
\rule[-1ex]{0pt}{3.5ex} 7 & White Light Fringes & ZPD Position, Fringe Quality\\ 
\hline
\rule[-1ex]{0pt}{3.5ex} 8 & Laser Fringes & Fringe Quality\\
\hline 
\end{tabular}
\end{center}
\end{table}

We have found, from application of this set of procedures over the
past six months, that the system does not appear to have any long-term
stability issues.  Because of the ongoing work on the system and
installation of new components, we have been unable to accurately
quantify the stability of individual elements, but as the final
components of the system are installed we will be able to precisely
determine our vulnerability to system drift.

\subsection{Camera Characterization}

\begin{figure}
   \begin{center}
   \begin{tabular}{c}
   \includegraphics[height=7cm]{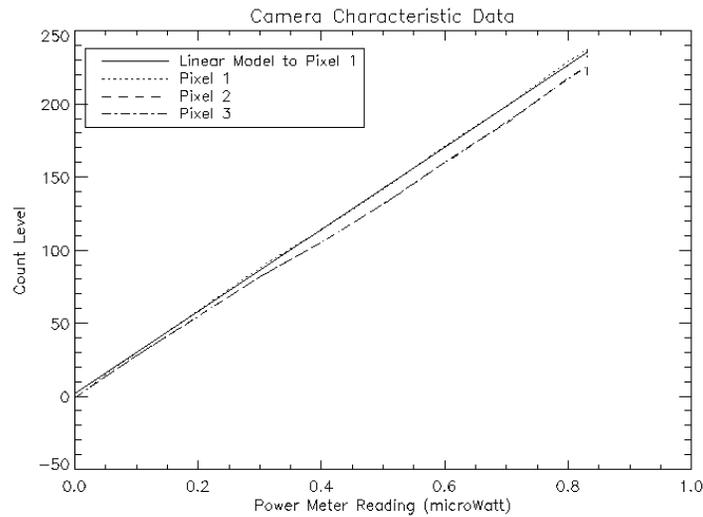}
   \end{tabular}
   \end{center}
   \caption[example] 
   { \label{fig:camchar} A plot of the count level of the pixels of the camera as a function of the amount of light in the beam.  We see that each of these pixels are remarkably linear for the full range of the detector.  There is some deviation at the brightest end of the scale, but this deviation is small and consistent across the detector, and can therefore be easily accounted for in software.}
   \end{figure}

In the process of testing and characterizing the system, we have
attempted to understand the behavior of all of the subsystems to the
greatest accuracy possible.  As an example of this effort, we include
here information on the characterization of the camera.  In specific,
the major topics of interest here are the linearity of response, dark
current and read noise, and uniformity of response across the array.
The dark current and read noise were tested by taking a number of dark
frames using different gains and integration times.  We found that the
dark current plus read noise occupied at most 3\% of the dynamic range
for all integration times of interest.  The uniformity of response was
measured using a series of flat field exposures.  From these, we found
that the individual pixel response was uniform to better than 1\%
where the major component of the nonuniformity here was due to
``striping'' on the detector due to slight offsets in black level and
gain for the two readout chanelns of the camera.  In order to measure
the linearity of response, we used a white light source at the focus
of the collimator with a secondary power monitor.  The power monitor
provides an accurate relative measure of light on the detector.  By
changing the amount of light, we are able to measure the response with
increasing source brightness.  A figure of the response is shown in
Figure \ref{fig:camchar}.  We find that the camera response is linear
for nearly the whole range.  Only at the highest end of the detector
range is there deviation from linearity, and it was found that this
deviation was of order 1\%, within the last 10\% of the camera's
dynamic range.  Further, we found that this behavior appeared to be
consistent for different pixels to very high accuracy.  We conclude
that for the majority of our observations, the camera can be assumed
to be completely linear.  For observations where the full dynamic
range of the camera is required, it may be necessary to correct for
the non-linearity in software, but given the uniformity of the
deviations across the array, this should be simple to implement.

%

\section{THE FUTURE OF WIIT: WIIT2} 
\label{sect:future}

WIIT is a valuable instrument which will be used to help prepare for
astronomical wide-field interferometry.  To continue improving the
system, several upgrades will be undertaken in the near future.
First, we will encase the interferometer inside rigid box, to provide
additional isolation from air turbulence and temperature variations
within the room.  Second, we will be installing a new Zygo
interferometer system with two seperate interferometer heads.  These
two measurement beams will be injected into the two arms of WIIT, so
that we are able to directly measure the optical path length of both
arms individually.  We will then be able to trigger not on the
position of the delay line stage, as done now, but on the actual
difference in distance for the two arms.  This will greatly improve
the system and provide another reduction in the effects from air
turbulence and temperature variations.  As a side note, it will also
be possible to use this to estimate the scale size of turbulence cells
around the apparatus and estimate overall sensitivity to air movement
around WIIT.  Third, we will improve the communications between the
individual computers.  This will improve the operability of the system
and will lead to the complete automation of data acquisition.
Finally, we will continue to characterize the system and understand
the limitations and abilities of individual components, including both
the 1-D components and the rotation stages for 2-D interferometry.

In addition to the work on WIIT itself, we will be using data from
WIIT to improve our understanding of algorithms and procedures for
dealing with interferometric data.  This will help us optimize data
acquisition for the system, and will be of direct bearing on the
methods used for future interferometry missions.


\acknowledgments     
 
We thank the WIIT Science and Technical Advisory Group (STAG) for
their sage advice and support.  The STAG members are Drs. Richard
Burg, Bill Danchi, Dan Gezari, Antoine Labeyrie, John Mather (Chair),
Harvey Moseley, Dave Mozurkewich, Peter Nisenson, Stan Ollendorf, Mike
Shao, and Hal Yorke.

Funding for WIIT is provided by NASA Headquarters through the
ROSS/SARA Program and by the Goddard Space Flight Center through its
IR\&D program.  S.A. Rinehart is funded through the NRC Research
Associateship Program.


\begin{center}
{\bf REFERENCES}
\end{center}

\noindent 1. D. T. Leisawitz, et al., ``Wide Field Imaging Interferometry Testbed I -- Purpose, Design and Data 

Products'', SPIE 4852-34, Waikoloa, August 2002.

\noindent 2. D. B. Leviton, et al., ``Wide Field Imaging Interferometry Testbed III -- Metrology Subsystem'', SPIE 

4852-99, Waikoloa, August 2002.

\end{document}